\documentclass[twocolumn,printnumbers,amsmath,amssymb,showpacs,prl]{revtex4}

\usepackage{graphicx}
\usepackage{color}

\begin{document}

\title{Shear induced solidification of athermal systems with weak attraction}

\date{\today}

\author{Wen Zheng}
\author{Hao Liu}
\author{Ning Xu$^{*}$}
\affiliation{CAS Key Laboratory of Soft Matter Chemistry, Hefei National Laboratory for Physical Sciences at the Microscale, and Department of Physics,
University of Science and Technology of China, Hefei 230026, People's Republic of China.}

\begin{abstract}

We find that unjammed packings of frictionless particles with rather weak attraction can always be driven into solid-like states by shear. The structure of shear-driven solids evolves continuously with packing fraction from gel-like to jamming-like, but is almost independent of the shear stress. In contrast, both the density of vibrational states (DOVS) and force network evolve progressively with the shear stress. There exists a packing fraction independent shear stress $\sigma_c$, at which the shear-driven solids are isostatic and have a flattened DOVS. Solid-like states induced by a shear stress greater than $\sigma_c$ possess properties of marginally jammed solids and are thus strictly-defined shear jammed states. Below $\sigma_c$, states at all packing fractions are under isostaticity and share common features in the DOVS and force network, although their structures can be rather different. Our study reveals the significance of the shear stress in determining properties of shear-driven solids and leads to an enriched jamming phase diagram for weakly attractive particles.

\end{abstract}

\maketitle

Particulate systems such as colloids, emulsions, foams, and granular materials can form disordered solids at high packing fractions \cite{liu1,liu2,hecke1,trappe1,ohern1,parisi1,bi1,lu1,lu2,poon1,zaccarelli1}.  The critical packing fraction of the transition from liquid-like to solid-like states is sensitive to the interaction \cite{bi1,lois1,song1} and geometry \cite{donev1,damasceno1,torquato1,schreck1} of particles. Consider the simplest case of static packings of spheres.  If the spheres are frictionless and purely repulsive, the transition happens as the jamming transition at a critical packing fraction $\phi_j$ \cite{ohern1}.  This jamming transition is signaled by the sudden formation of a rigid and isostatic force network, i.e., the average coordination number is equal to twice of the dimension of space.  If the spheres are frictional, jamming can happen at lower packing fractions \cite{somfai1,song1}.  Interestingly, originally unjammed packings of frictional spheres below $\phi_j$ can jam under quasistatic shear, which is called shear jamming \cite{bi1}.

Shear jamming can occur at $\phi<\phi_j$ because there exist jammed states of frictional particles, while the shear just provides an opportunity to search for them in the sea of unjammed states.  Analogically, attraction may play a similar role as friction in helping produce jammed islands in the unjammed sea below $\phi_j$. It is then interesting to know whether shear forces could effectively drive unjammed states into solid-like states in the presence of attraction, especially in the zero attraction limit.

Attraction is however qualitatively different from friction, so are the induced phase behaviors. Compared with packings of frictional spheres which can only jam above a lower packing fraction limit \cite{bi1,somfai1,song1}, attraction can induce more complicated solid-like phases over a wide range of packing fractions, e.g., gels and glasses at low and high packing fractions \cite{poon1,zaccarelli1,lu1,lu2,lois1}.  If shear forces help to form solids in the presence of attraction, we may confront various types of solids and have to tackle how to distinguish them in terms of not only packing fraction but also shear stress. The output is significant to enriching our knowledge of the shear-driven transition from liquid-like to solid-like states and completing the jamming phase diagram for attractive particles \cite{trappe1}.

In this letter, we investigate how attraction affects the picture of the jamming of purely repulsive systems \cite{ohern1} at zero temperature ($T=0$) and finite shear stresses.  We study the formation and properties of shear-driven solids of particles interacting via a repulsive core and a tiny attractive tail. By applying quasistatic shear, we can always find solid-like states below the jamming transition, where the probability of finding solid-like states is almost zero in the absence of shear. By minimizing a thermodynamic-like potential \cite{liuhao1}, we efficiently sample solid-like states at fixed packing fraction $\phi$ and shear stress $\sigma$, and analyze their structure, force network, and vibrational properties. The analysis enables us to construct an extended jamming phase diagram in the $\sigma-\phi$ plane at $T=0$. In particular, there is a packing fraction independent shear stress $\sigma_c$. When $\sigma=\sigma_c$, shear-driven solids exhibit features of marginally jammed solids, e.g., being isostatic and having a flattened density of vibrational states (DOVS) \cite{ohern1,silbert1,wyart1}. Solid-like states driven by $\sigma>\sigma_c$ are thus strictly-defined shear jammed states.

Our systems are two dimensional with side length $L$ in both directions. To avoid crystallization, we put $N/2$ large and $N/2$ small disks with equal mass $m$ in the system.  The diameter ratio of the large to small particles is $1.4$. Here, we show results for $N=1024$ systems. The interparticle potential is \cite{lois1,irani1}
\begin{equation}
U(r_{ij})=
\begin{cases}
\frac{\epsilon}{2}\left[\left( 1-\frac{r_{ij}}{d_{ij}}\right)^2-2\mu^2\right],\frac{r_{ij}}{d_{ij}}\le 1+\mu,\\
-\frac{\epsilon}{2}\left( 1+2\mu-\frac{r_{ij}}{d_{ij}}\right)^2,\ \ 1+\mu<\frac{r_{ij}}{d_{ij}}\le 1+2\mu,\\
0,\ \ \ \ \ \ \ \ \ \ \ \ \ \ \ \ \ \ \ \ \ \ \ \ \ \ \ \frac{r_{ij}}{d_{ij}}>1+2\mu,\label{potential}
\end{cases}
\end{equation}
where $r_{ij}$ and $d_{ij}$ are the separation between particles $i$ and $j$ and sum of their radii, and $\mu$ is a tunable parameter to control the range and strength of attraction.  We vary $\mu$ from $10^{-2}$ to $10^{-6}$, approaching the zero attraction limit. When $\mu=0$, the harmonic repulsion widely employed in the study of jamming \cite{ohern1} is recovered. We set the units of mass, energy, and length to be particle mass $m$, characteristic energy scale of the potential $\epsilon$, and small particle diameter $d_s$.

\begin{figure}[t]
\vspace{-0 in}
\includegraphics[width=0.48\textwidth]{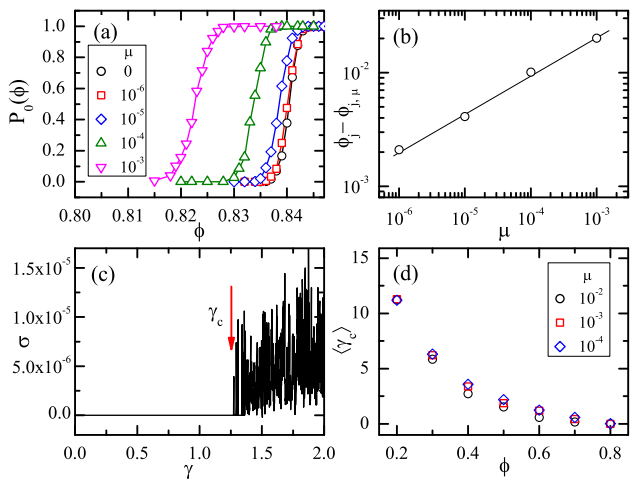}
\vspace{-0.25 in}
\caption{\label{fig:fig1} (color online). (a) Probability of finding jammed states without shear ($\gamma=0$), $P_0(\phi)$, for different strength of attraction $\mu$. The lines are to guide the eye. (b) Critical scaling of the jamming transition threshold: $\phi_j-\phi_{j,\mu}\sim \mu^{1/3}$, with the line having a slope of $1/3$. (c) Stress-strain relation of an initially unjammed state at $\phi=0.60$ and $\mu=10^{-3}$ under quasistatic shear. The arrow points to the onset shear strain $\gamma_c$ at which solid-like states start to be explored. (d) Packing fraction dependence of $\left<\gamma_c\right>$ for different $\mu$.
}
\end{figure}

The shear deformation is realized by introducing the shear strain $\gamma$ and applying the Lees-Edwards boundary conditions \cite{allen_book}. Without shear ($\gamma=0$ or remains constant), we generate $10000$ static states at fixed packing fraction by applying the fast inertial relaxation engine minimization method \cite{fire} to minimize the potential energy $U=\sum_{ij}U(r_{ij})$ of initially random configurations, where the sum is over all pairs of interacting particles. Figure~\ref{fig:fig1}(a) shows the probability $P_0(\phi,\mu)$ of finding solid-like states ($|U|/N > 10^{-16}$) at $\gamma=0$. With increasing $\mu$, $P_0(\phi,\mu)$ shifts to lower packing fractions. Employing the definition in Ref.~\cite{ohern1}, we determine the jamming transition threshold $\phi_{j,\mu}$. As shown in Fig.~\ref{fig:fig1}(b), $\phi_j - \phi_{j,\mu}\sim \mu^{1/3}$ in the small $\mu$ limit, where $\phi_j\approx 0.842$.

Figure~\ref{fig:fig1}(a) indicates that well below $\phi_{j,\mu}$ the direct-quench sampling at $\gamma=0$ is almost impossible to find solid-like states. Starting from unjammed states, we successively increase $\gamma$ by a step size $\Delta \gamma$, followed by a potential energy minimization. As shown in Fig.~\ref{fig:fig1}(c), in early stage of this quasistatic shear, the system remains unjammed ($\sigma=0$). Interestingly, as long as $\Delta \gamma$ is small enough, nonzero shear stress emerges and fluctuates when $\gamma>\gamma_c$, signaling the formation of solid-like states. The solid forming ability under quasistatic shear decreases with decreasing packing fraction, demonstrated by the growth of $\left<\gamma_c\right>$ in Fig.~\ref{fig:fig1}(d), where $\left< .\right>$ denotes the average over independent runs of quasistatic shear.

The shear stress fluctuates and is not controllable during quasistatic shear. To have a clear picture of the shear stress dependence (which turns out to be important), we instead sample shear-driven solids by a recently developed algorithm \cite{liuhao1} to well control the shear stress. By minimizing a thermodynamic-like potential $H=U - \sigma\gamma L^2$ of arbitrary configurations, we can quickly look for solid-like states at desired shear stress $\sigma<\sigma_y$, where $\sigma_y$ is the yield stress, i.e., the maximum shear stress for shear-driven solids to exist. As shown in Fig.~\ref{fig:fig1}, it takes some shear strain to find shear-driven solids. Furthermore, it is impractical to let $\gamma\rightarrow \infty$. We thus set a maximum strain $\gamma_m=20$. The search for shear-driven solids fails once $\gamma>\gamma_m$.

\begin{figure*}[t]
\vspace{-0.2 in}
\includegraphics[width=0.9\textwidth]{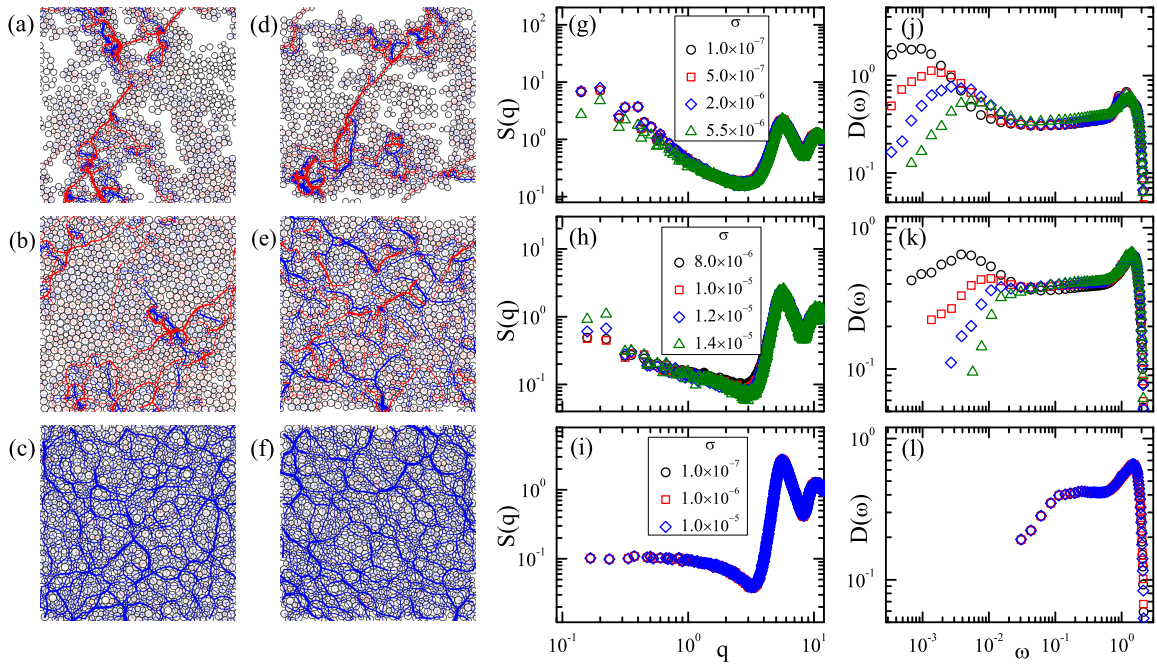}
\vspace{-0.15 in}
\caption{\label{fig:fig2} (color online). (a)-(f) Snapshots, (g)-(i) static structure factor $S(q)$, and (j)-(l) DOVS $D(\omega)$ for solid-like states ($\mu=10^{-3}$) driven by different shear stresses and at different packing fractions: $\phi=0.60$ (top row), $0.75$ (middle row), and $0.84$ (bottom row). For (a)-(c), $\sigma=10^{-6}$ (well below the yield stress), while for (d)-(f) $\sigma=6\times10^{-6}$, $2\times10^{-5}$, and $2\times10^{-4}$, respectively, which are all near the yield stress. The red (blue) lines in (a)-(f) are attractive (repulsive) interactions, with the thickness illustrating the strength. To have a better vision, we normalize the interaction strength by the maximum value for every snapshot. The shear stress values of the $D(\omega)$ curves are listed in the legend of the $S(q)$ panels to the left.
}
\end{figure*}

\begin{figure}[t]
\vspace{-0 in}
\includegraphics[width=0.48\textwidth]{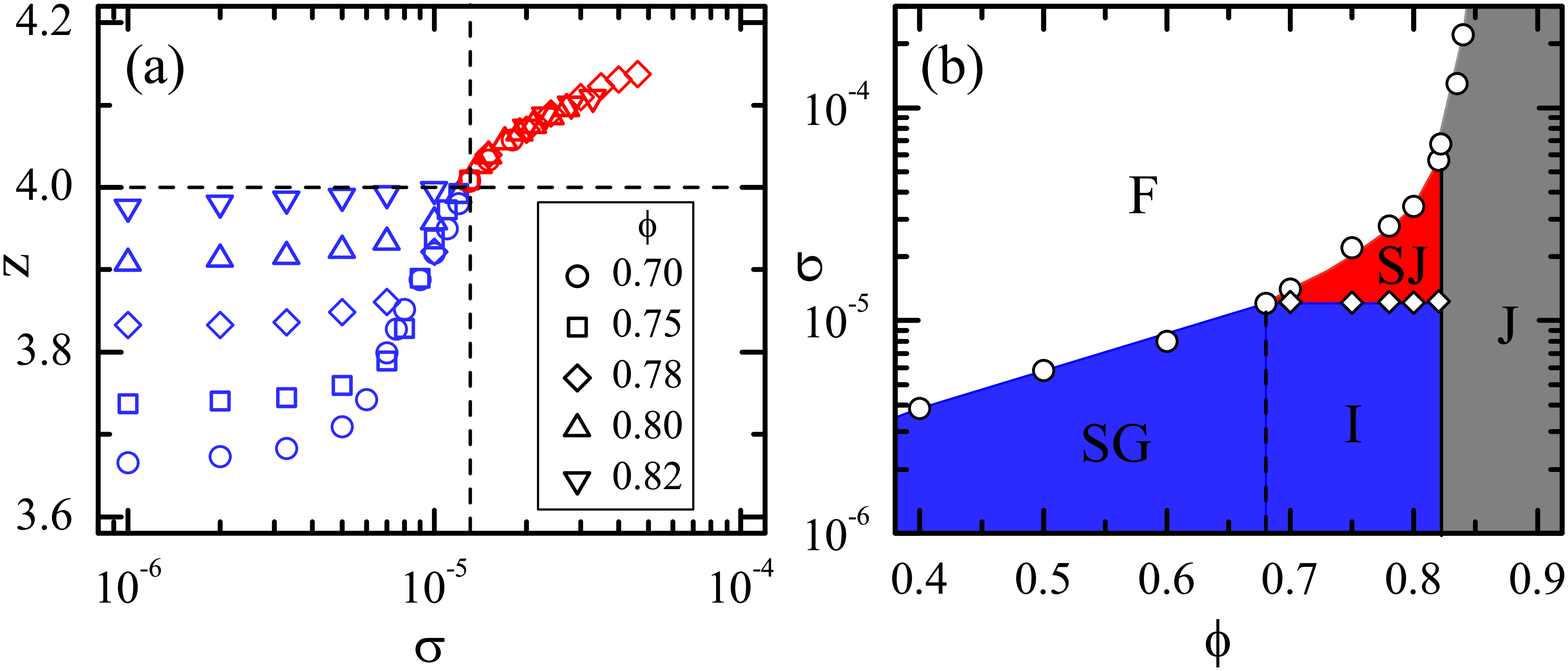}
\vspace{-0.25 in}
\caption{\label{fig:fig3} (color online). (a) Shear stress dependence of the average coordination number $z(\sigma)$ for shear-driven solids ($\mu=10^{-3}$) at $\phi<\phi_{j,\mu}$. At each packing fraction, the blue (red) symbols denote data below (above) the crossover shear stress $\sigma_c$ and with (without) low-frequency peak in the DOVS. The horizontal and vertical dashed lines label $z=4$ and $\sigma_c\approx 1.2\times 10^{-5}$. (b) Extended jamming phase diagram for weakly attractive particles ($\mu=10^{-3}$) at $T=0$. The circles and diamonds are the yield stress $\sigma_y(\phi)$ and crossover shear stress $\sigma_c(\phi)$, respectively. The solid and dashed vertical lines are $\phi=\phi_{j,\mu}$ and $\phi=\phi_{rp}$. F, SJ, SG, I, and J denote regimes of flowing (white), shear jammed (red), shear gel-like (blue, $\phi<\phi_{rp}$), intermediate (blue, $\phi_{rp}<\phi<\phi_{j,\mu}$), and jammed (gray, $\phi>\phi_{j,\mu}$) states, respectively.
}
\end{figure}

For each pair of $\phi$ and $\sigma$, we run $1000$ independent trials. When $\sigma$ is small, we can always find solid-like states. The probability of successful trials decreases with increasing $\sigma$ near $\sigma_y$. When $\sigma>\sigma_y$, the search for shear-driven solids always fails. In Fig.~\ref{fig:fig3}(b) of the extended jamming phase diagram, we show an example of $\sigma_y(\phi)$ for $\mu=10^{-3}$. $\sigma_y$ decreases with decreasing $\phi$, and remains nonzero down to rather low packing fractions. At fixed $\phi$, $\sigma_y$ decreases when $\mu$ decreases. In particular, we find that $\sigma_y\sim \mu$ at low packing fractions. Therefore, in the small attraction limit, attraction does not act as a perturbation \cite{berthier3,zhang1,wang3}, whereas it always induces multiple types of solid-like states far below the jamming threshold and qualitatively alters the jamming phase diagram for purely repulsive particles.

Systems at $\phi>\phi_{j,\mu}$ are essentially jammed without the need of shear, which do not interest us here. The focus of this work is on the regime of $\phi<\phi_{j,\mu}$, where direct quenching always finds unjammed states and solid-like states can be explored with the help of shear stresses smaller than $\sigma_y$. In the following, we will mainly discuss shear-driven solids at $\phi<\phi_{j,\mu}$. Results for $\phi>\phi_{j,\mu}$ are presented just for comparison.

Figures~\ref{fig:fig2}(a)-(f) are configurations with force network of the shear-driven solids at different packing fractions and shear stresses. At $\phi\ll \phi_{j,\mu}$, attraction (red bonds) dominates and the states look gel-like with fractal structures. Slightly above $\phi_{j,\mu}$, the structure looks uniform and particle interactions are predominantly repulsive (blue bonds). In between, with increasing packing fraction, the structure evolves from gel-like to jamming-like.

Figures~\ref{fig:fig2}(g)-(i) demonstrate the packing fraction and shear stress evolution of the static structure factor $S(q)=\left<|\sum_j{\rm exp}(i\vec{q}\cdot \vec{r}_j)|^2\right>/N$, where $q=|\vec{q}|$ is the angular wavenumber, $\vec{r}_j$ is the location of particle $j$, $\left< .\right>$ denotes the average over solid-like states, and the sum is over all particles. The structure is almost independent of shear stress, while it evolves strongly with packing fraction. At $\phi\ll \phi_{j,\mu}$, the low $q$ part of $S(q)$ exhibits the typical gel-like feature, $S(q)\sim q^{-d_f}$ with $d_f\le 2$ being the fractal dimension \cite{tanaka1}. The low $q$ part of $S(q)$ moves down with increasing packing fraction, and eventually becomes flat (jamming-like feature \cite{donev2,xu2,berthier2}) near $\phi_{j,\mu}$.

Purely from the packing fraction evolution of $S(q)$, we cannot determine the boundary between gel-like and glass or jamming-like states. Note that the solid-like states are shear induced. Although the structure is insensitive to the change of shear stress, other quantities may exhibit shear stress dependence and provide useful information to distinguish states. Comparing states at the same packing fraction but different shear stresses [e.g., Figs.~\ref{fig:fig2}(b) and \ref{fig:fig2}(e)], we can tell that the shear stress indeed remarkably affects the force network: More particles interact and repulsion plays a more important role with increasing shear stress.

Resulting from significant changes in the force network, vibrational properties of shear-driven solids exhibit strong shear stress dependence.  Figures~\ref{fig:fig2}(j)-(l) show the shear stress evolution of the DOVS, $D(\omega)=\left< \sum_l \delta(\omega - \omega_l)\right> / 3N$, where $\omega_l$ is the frequency of the $l^{th}$ normal mode of vibration, $\left< .\right>$ denotes the average over configurations, and the sum is over all modes. The normal modes of vibration are obtained from diagonalizing the Hessian matrix using ARPACK \cite{arpack}. When $\phi>\phi_{j,\mu}$, applying shear stress only weakly affects the force network and elastic properties \cite{liuhao1}. As shown in Fig.~\ref{fig:fig2}(l), $D(\omega)$ at different shear stresses overlap. Interestingly, Figs.~\ref{fig:fig2}(j) and (k) show that $D(\omega)$ has strong shear stress dependence when $\phi<\phi_{j,\mu}$. For solid-like states induced by small shear stresses, there is a low-frequency peak in $D(\omega)$, indicating the aggregation of soft modes. With increasing shear stress, the peak moves down and to higher frequencies, implying the decrease of the amount of soft modes and that shear-driven solids become stiffer and more stable.

At all packing fractions below $\phi_{j,\mu}$, the motion of the low-frequency peak in $D(\omega)$ with the change of shear stress follows the same trend. However, at low packing fractions where shear-driven solids are typically gel-like, until at the yield stress, the peak is still present. In contrast, near $\phi_{j,\mu}$, the peak disappears at a crossover shear stress $\sigma_c<\sigma_y$. Meanwhile, $D(\omega)$ exhibits a plateau, which is actually one of the most representative features of marginally jammed solids of purely repulsive particles \cite{silbert1,wyart1,wang1}. When $\sigma>\sigma_c$, the evolution of $D(\omega)$ looks like that of marginally jammed solids under compression, but here the shear stress is the driving force instead of the packing fraction.

For marginally jammed solids, the flattening of $D(\omega)$ is associated with isostaticity \cite{silbert1,wyart1,wang1}, i.e., the average coordination number $z=z_c=2d$ with $d$ being the dimension of space. Is the emergence of the plateau in $D(\omega)$ at $\sigma_c$ also related to isostaticity?

Figure~\ref{fig:fig3}(a) shows the shear stress evolution of the average coordination number (rattlers excluded) at $\phi<\phi_{j,\mu}$. When the shear stress is small, there is a plateau in $z(\sigma)$. The plateau value approaches $z_c$ from below with increasing packing fraction. Approaching the yield stress, the coordination number grows quickly and collapses onto a master curve. In each $z(\sigma)$ curve, data points at $\sigma<\sigma_c$ and with the low-frequency peak in $D(\omega)$ are denoted by blue symbols, while red symbols represent data at $\sigma>\sigma_c$. Surprisingly, $z=z_c=4$ is exactly the boundary between two colors, indicating that isostaticity is indeed coupled to the flattening of $D(\omega)$ at $\sigma_c$. Moreover, the collapse of all data at $z>z_c$ implies that the value of $\sigma_c$ is independent of the packing fraction. We estimate $\sigma_c$ at various packing fractions and find that it is indeed constant in $\phi$, as shown in Fig.~\ref{fig:fig3}(b).

Now we see that solid-like states driven by a shear stress greater than $\sigma_c$ possess important features of marginally jammed solids, such as $z>z_c$ and jamming-like $D(\omega)$. It is thus plausible to strictly define them as shear jammed solids.

Our major findings lead to the extended jamming phase diagram for weakly attractive particles in the $\sigma-\phi$ plane at $T=0$. Figure~\ref{fig:fig3}(b) is an example of the diagram for $\mu=10^{-3}$.

Strictly-defined shear jamming is encircled by $\phi=\phi_{j,\mu}$, $\sigma_y(\phi)$, and $\sigma=\sigma_c$. Interestingly, $\sigma=\sigma_c$ intersects $\sigma_y(\phi)$ roughly at the rigidity percolation threshold, $\phi_{rp}\approx 0.689$ \cite{lois1}. We have verified for other values of $\mu$ that this is not a coincidence. As mentioned earlier, solely from $S(q)$, it is hard to determine the crossover packing fraction to separate gel-like states from jamming-like states. Now that shear jammed states only exist at $\phi>\phi_{rp}$, $\phi=\phi_{rp}$ is a plausible candidate of such a crossover.

Shear-driven solids lying below $\sigma=\sigma_c$ share some common features, e.g., existence of the low-frequency peak in $D(\omega)$, $z<z_c$, and attraction dominant, although they cover a wide range of packing fractions and exhibit progressive packing fraction evolution of the structure. States between $\phi_{rp}$ and $\phi_{j,\mu}$ are particularly interesting. They have similar structures to shear jammed states but resemble shear gel-like states at $\phi<\phi_{rp}$ in mechanical and vibrational properties. We tentatively name them as intermediate states. The existence of intermediate states can only be found by the careful study of the shear stress dependence. It also warns us about the danger to identify various types of amorphous solids from structure \cite{tanaka1} or vibrational properties \cite{lohr1} alone.

In summary, in the presence of weak attraction, athermal solid-like states are explored by shear over a wide range of packing fractions below the jamming transition. Our careful study of the packing fraction and shear stress dependence reveals that the static structure of shear-driven solids is sensitive to the change of packing fraction, but not to shear stress. In contrast, the DOVS and force network evolve progressively with shear stress. The strong shear stress dependence enables us to determine strictly-defined shear jamming and construct an extended jamming phase diagram in the $\sigma-\phi$ plane.

As shear stress increases, the rigidity of shear-driven solids at $\phi<\phi_{j,\mu}$ increases, reflected in the decay of soft modes and increase of the coordination number and elastic moduli (not shown). In contrast, increasing shear stress slightly softens jammed solids well above $\phi_{j,\mu}$. This opposite behavior on both sides of $\phi_{j,\mu}$ is analogous to that of thermal systems: With increasing temperature, glasses are hardened below the jamming-like transition, while slightly softened above \cite{ikeda1,wang2}. Therefore, we provide evidence supporting that the shear stress can have similar effects as the temperature on transitions between liquid-like and solid-like states and properties of amorphous solids \cite{lee1}, as proposed by the original jamming phase diagram \cite{liu1}.

Our work is relevant to experimental systems like granular materials and non-Brownian colloids. For colloidal systems with Brownian motion, how temperature affects shear induced solidification is interesting to attack next. Both the thermal motion and shear can harden systems at $\phi<\phi_{j,\mu}$. It is quite interesting to figure out whether and how they may compete or help each other to induce unpredictable results in dynamics and phase behaviors.

We are grateful to Ke Chen and Jie Zhang for helpful discussions.  This work is supported by National Natural Science Foundation of China No.~21325418 and 11574278, National Basic Research Program of China (973 Program) No.~2012CB821500, and Fundamental Research Funds for the Central Universities No.~2030020028 and 2030020023. We also thank the Supercomputing Center of University of Science and Technology of China for computer times.

\end{document}